# Distribution of U and Th and Their Nuclear Fission in the Outer Core of the Earth and Their effects on the Geodynamics

Xuezhao Bao

CFI high pressure and high temperature lab, Department of Earth Sciences, the University of Western Ontario, Lodon, Ontario, Canada, N6A 5B7
Email: xbao2@uwo.ca
xuezhaobao@hotmail.com

**Abstract:**

Here we propose that there is a lot of heat producing elements U and Th in the Earth's outer core. The heat released from them may be the major energy source for driving the material movement within Earth's interior, including plate motion. According to seismic tomography, the hottest area is the mantle under the central Pacific Ocean. Combined with geomagnetic data, it is derived that the magnetic and heat convection centers deviate from Earth's geographic center to the Pacific direction for 400 km. Therefore, U and Th may be more concentrated in a position close to the equator in the lower outer core (from the middle to bottom of the outer core) under the central Pacific Ocean, and have formed a large U, Th-rich center there. Another small U, Th-rich center may be located in a position close to the equator in the lower outer core under Africa, which is directly opposite of the large U, Th-rich center past the solid inner core. The two U, Th-rich centers may have led to the formation of the Pacific and Africa super-plumes and are offering energy to run the plate tectonic system. The large U, Th-rich center also could have caused the temperature of the western hemisphere to be higher than that of the eastern hemisphere in the inner core, which may be the cause for the east-west hemispherical elastic anisotropy of the inner core. Periodical nuclear fissions of U and Th may have occurred in the outer core in Earth's geological history. This kind of natural fission of U and Th may be similar to the spontaneous fissions that occurred in Precambrian uranium mines. This is also supported by planetary quality and luminosity relation studies. Another piece of evidence comes from noble gases from volcanic rocks and their xenoliths in Pacific volcanic hot spots; since some of these noble gases could only be produced by nuclear fission of U and Th. Periodical U, Th fissions in the outer core might have triggered geomagnetic superchrons and reversals. At the same time, the energy released from the outer core during these events might have also triggered strong and extensive global geological and volcanic activities, and caused mass extinctions on the surface. It may have also led to Earth's cyclical expansions in its geological history.

*Key words:* U, Th distribution in the outer core; nuclear fission; noble gases; geodynamics; geomagnetic superchrons and reversals; elastic anisotropy of the inner core.

## 1. Introduction

The energy source that drives material movement within Earth's interior remains controversial [1~3]. Plate tectonics theory successfully explains various geological phenomena in the interior and edge regions of oceans, and is trying to explain continental dynamics [4], but it fails to elucidate the driving force behind plate motion [1~3]. Some scholars proposed that the force produced by the differential rotation of different layers (crust, mantle, core, etc.) of the Earth [2], or the force induced by other celestial bodies are the driving forces for plate movement. However, these theories cannot explain why plate tectonics only exists on Earth, but not on other terrestrial planets, such as Mars, Venus and Mercury [5,6], since these forces or energy sources should also exist in them (a more reasonable explanation can be found in: Terrestrial planetary dynamics: a view from U, Th geochemistry. At http://arxiv.org, arXiv: astro-ph/0606468, June 2006). Therefore, we must study Earth's interior to find the driving force for plate motion.

Among Earth's materials, only radioactive elements are energy-producing. The driving force for moving materials must come from a source of energy. Therefore, understanding the distribution of radioactive elements in Earth's interior may play a key role in understanding geodynamic problems. In the early 1970s, Wilson [7] and Morgan [8] recognized that the driving force of plate motion may be related to the mantle plumes, which originate from the deep mantle, namely, this force may originate from the lower mantle, even from the core-mantle boundary (CMB). Modern computer simulations and seismic tomographic mapping techniques also show that the driving force causing mantle convection must come from the Earth's core [9, 10]. This is contradictory to traditional geochemical models. Most of these models predict that most radioactive elements reside in the Earth's crust [2, 11~14] and upper mantle, and they are negligible in the core [2]. Some researchers speculated that heat contribution generated by radioactive elements is 76% from the crust, 21% from the mantle, and 2.0% from the core [11~14].

In order to determine what the driving force is, many researchers proposed that when material from Earth's crust is subducted to the deep mantle, the heat released from U and Th in the subducted crust materials could be the energy source to drive mantle convection. But Griffiths concluded after calculations that this is impossible [15]. Our previous work also shows that U and Th gradually migrate from Earth's deep interior to the continental crust, and these elements cannot return to Earth's interior in significant amounts due to their large ionic diameter and strong affinity for O and other oxidative volatiles, including $H_2O$ [5, 6]. U and Th usually remain in the crust, especially the continental crust after migrating from the deep interior. This is a way for U and Th to accumulate in the crusts (mainly continental crust) with geological time [5, 6, 13, 14]. Anderson suggested that in the bottom of the mantle, there is a ~ 200-km layer being enriched with CaO-A1$_2$O3-TiO$_2$, etc. refractory (**CaO-A1$_2$O3-TiO$_2$-rich**) materials. These refractory materials may contain a high content of U and Th. These CaO-A1$_2$O$_3$-TiO$_2$-rich materials with high melting points might have been formed during Earth's accretion and remains as residues in the bottom of the mantle. Since the density of these CaO-A1$_2$O$_3$-TiO$_2$-rich materials is much smaller than that of the ferromagnesian mantle, they will rise with U and Th, and the heat released from U and Th will trigger the formation of mantle plumes [16]. Anderson's chemical plume hypothesis still is problematic: firstly, the Earth has evolved for 4.5 Ga; therefore, these CaO-$Al_2O_3$-$TiO_2$-rich light materials driven by the heat from U and Th should have been moved to the crust in Earth's early stages. Paleomagnetic studies have shown that catastrophic geological activities in the crust always followed the end of a geomagnetic superchron and strong variation in geomagnetic field in a short time interval [17~20] (note: currently this interval is considered being in the range of 10-20 Ma(Courtillot and Olson, EPSL 260, 2007 p495)). This means that the strong energy variation in the outer core can be transferred to the crust in a short geological time interval, since the magnetic field is thought of as having originated from the convection in the outer core (see below). Therefore, these materials containing heat producing elements with a small density are able to rise quickly to the upper mantle and crust. They cannot be the source of the long-term driving force for plate motion. Secondly, it cannot explain why plate tectonics only exists on Earth and not on other territorial planets. According to a heterogeneous accretion theory of planet formation, other planets should also have similar CaO-$Al_2O_3$-$TiO_2$-rich materials with a high content of U and Th, and similar chemical plumes in their interiors. For instance, many similar heat mantle plumes were found in Venus [21, 22]. However, a corresponding plate tectonics system has not been found [22~24]. Other researchers, such as Davies and Schubert, believed that the heat from radioactive elements can maintain mantle convection, but they did not give an explanation about the distribution of these radioactive elements [25]. Therefore, the distribution of heat producing elements within Earth's interior is still unclear [12-13].

In addition, the findings of the Earth's compositional heterogeneity do not support the assumption that "original building materials of Earth consist of uniform chondrite meteorites" [26] (new studies also conclude that the original building materials of the Earth are unlike any known chondrite group or mixture of chondrite group (Righter, 2003. Annu. Rev. Earth Planet. Sci. 32 p135)). Therefore, the total amount of U and Th in the Earth's interior calculated from the chondrite meteorite compositional model also is also problematic. Consequently, it is necessary to re-evaluate the U and Th distribution model [2] and the total amount of U and Th within Earth's interior to understand geodynamics. In our recent work, a new U, Th geochemical model has been proposed [5,6]. This new model suggests that the reducing conditions in the Earth's core and lower mantle have led some U and Th to migrate to the Earth's core [5,6]. Currently, much of these U and Th are still in the outer core [5,6]. This conclusion is supported by the experimental partitioning of U and Th between silicates and metal sulfides [27] (please also refer to our new experimental results (Bao, et al., 2006. http://arxiv.org, astro-ph/0606614, June 2006; Bao and Secco, 2006. http://arxiv.org, astroph/0606634, June 2006; Bao, 2006. http://arxiv.org, astro-ph/0606468, June 2006)). Planetary quality and luminosity relation studies also indicate of existing nuclear reactions in Earth's interior [28]. Earth's interior never reaches a temperature of $1x10^9$ $^oC$, which is required for nuclear fusion reaction [29]. Therefore, this kind of nuclear reaction may be U and Th nuclear fission [30, 31]. Rocks coming from the mantle are usually rich in noble gases produced by U, Th decay or fission, which also supports such nuclear fission reactions in Earth's interior [31].

The assumption that there is a lot of U and Th in Earth's outer core makes it easier to understand Earth's evolution in many aspects, and it also provides insight into the dependence of life evolution on the Earth's evolution [5,6]. This article will further discuss U and Th in the outer core and their relationship with geodynamics, with arguments from geo-magnetic and seismic physics, and supportive data on Earth's interior temperature and isotope composition of noble gases, etc.

## 2. The geochemistry of U and Th

The following are the basic points of U, Th geochemistry [5,6]: 1) of all naturally occurring elements, U and Th have the biggest density, and high melting and boiling points. As a result, under highly reducing conditions, they will exist as stable metals or low valence oxides and compounds; they cannot migrate upwards, and will sink into the core. 2) At the same time, U and Th have large atomic radii, and lose their valence electrons readily. Therefore, they have a strong tendency to be oxidized by oxidative volatile elements or their ions ($O_2$,$H_2O$, $N_2$, S, C etc.) to form components and complexes with a low melting point, and small density [5,6]. These components and complexes have a strong tendency to migrate up to the crust.

The two opposite migration tendencies of U and Th have led to different U, Th migration patterns between Earth's upper and lower parts. The upper part of the Earth: the U and Th in the rock will gradually be oxidized by $O_2$, $H_2O$, $F_2$, $N_2$, and other oxidative volatile elements or their ions, and migrate upwards to the crust. In the first stage, they will be moved to the asthenosphere position, and are enriched there to form an enrichment zone of U and Th (EZ) under the lithosphere [5-6]. Then, mantle fluid [32], and oxidative volatile components or their ions brought down by subducting plates further oxidize U and Th in the EZ to form high valence components and complexes with a lower melting point and lower density, and finally they are moved up to the crust [5-6].

It should be pointed out that the lower part of the lithosphere that consists of melt-resistant (refractory) ultramafic rocks can act as thermal and composition insulation layers, which prevents the heat released from U and Th in the EZ to rise to the surface, and also prevents water and volatile components from the surface to enter the EZ. Namely, rocks in the EZ cannot exchange materials or heat with the Earth's surface. Therefore, most of the rocks in the EZ are under highly reducing conditions. Consequently, U and Th in the EZ still exist as low valence oxides and compounds with a high melting point and large density, and do not migrate with magma up to the crust. Since oxidative volatile components cannot be moved down to the EZ in the mid-ocean ridge positions, therefore, U and Th in the EZ under these positions cannot combine with enough water and oxidative volatile components to migrate up to the crust. Hence, the concentrations of U and Th are very low in magma erupted from mid-ocean ridges. However, in the subduction zones, rocks in the subducing oceanic plates are rich in water and other volatile components [33]. Therefore, water and oxidative volatile components can be brought down to the EZ with these rocks. Consequently, low valence oxides and components of U and Th in the EZ under these positions will be further oxidized by water and oxidative volatile components to form mobile high valence U and Th components and complexes, and finally migrate up to the crusts with magma. For instance, U and Th concentrations are relatively high in the rocks from the Circum-Pacific volcanic belts [34]. This is because these rocks are produced by the subduction in the Circum-Pacific subduction zones. The volatile-rich kimberlites coming from the deep mantle are also rich in U and Th[35]. This indicates that water and oxidative volatile components can oxidize the stable U and Th oxides and compounds in the mantle, and help them to migrate up to the crust through magmatism.

The lower mantle and core are extremely reducing and lack volatile components. U and Th exist as metals or low valence compounds and oxides there. U and Th metal, and their low valence compounds and oxides have high densities, and melting and boiling points. Consequently, they are stable, and cannot migrate upwards. On the contrary, when the surrounding rocks are in a state of melting or partially melting, U, Th will sink into the core. Experimental studies have shown that under highly reducing conditions, U and Th will gradually change from lithophile to sulfophile character; therefore, they enter the core more easily than K and become an important energy source in planetary cores [27]. Currently, these U and Th are still partially in the outer core [5-6]. Since the lower mantle and core account for a major part of Earth, the total amount of U and Th in the outer core may be significant.

# 3. The relationship between U, Th in the core and geodynamics

## 3.1 Energy and U, Th uneven distribution in the outer core

### 3.1.1 Energy in the outer core

Since 4.5 Ga ago, Earth has been releasing its internal heat energy through volcanic and other geological activities, and to this day its huge outer core is still molten with extremely high temperature (4000 k -5000 k). Therefore, the amount of energy in Earth's interior, particularly in the core, should be enormous. In addition to the above geochemical arguments, the following will indicate the existence of a large amount of energy in the core.

(1) The Geomagnetic field is thought of as having been produced by thermal convection in the liquid outer core [2, 17-20, 36], and it has been working this way for at least 1.0 Ga. To maintain this gigantic magnetic field, the core must have sufficient energy to support it [2].

(2) In accordance with the dynamo theory [2, 17~20, 36], the formation of a dipolar magnetic field requires differential rotation between the conductive fluid in the outer core and the solid mantle on its top. According to Ditfurth, if there is no other source of energy in the core, the differential rotation will stop soon. For example, a cup of tea in rotation will have consistent motion between tea leaves, water and cup after a while [36].

(3) Recent studies show that during the superchrons of the geomagnetic fields (a period without geomagnetic pole reversal), sea-floor spreading, global climate variation, plume activities, and extensive magmatic activities were exceptionally strong [2, 17~20]. At the same time, mass extinctions, including the extinction of dinosaurs, occurred [17]. Furthermore, these events have occurred periodically [2, 17~20]. Since the geomagnetic field is produced through convection in the liquid outer core, therefore, the strong geological activities in the crust are essentially triggered by strong energy release in the outer core. According to traditional radioactive element abundance and distribution patterns, which assume that U and Th reside mainly in the crust, with some in the upper mantle [11~14], it is difficult to explain the cyclical explosions of energy from the outer core. However, considering that U and Th have entered the core, and they have repetitive nuclear fissions, then this issue can be reasonably explained (see 3.2.2 and part 4).

(4) Since O.C. Hilgenberg (1933) proposed a hypothesis on the expansion of the Earth, it has gotten support from some scholars. Wang Hongzhen also suggested a hypothesis of Earth's cyclical asymmetric limited expansion [3]. He proposed that there were five expansion phases with an interval in ~500 Ma at super-continental periods around 2500Ma, 1950Ma, 1450Ma, 850Ma and 250Ma. The radius of the Earth during these phases is 75%, 80%, 85%, 90% and 95% of the present size respectively. That is, each expansion event increased Earth's radius by 5% [3]. This indicates that during these expansion phases, the temperatures in earth's interior might have increased rapidly. The studies of many researchers, including HC Urey, AE Ringwood and DL Anderson show that Earth's interior has been heating up [2]. Since the Earth has been releasing its internal heat through volcanic activities etc., it was possible for it to maintain a continuous warm-up in its interior only when its total heat-producing elements increased over time. In fact, because of the decay of radioactive elements, this total has been declining [5,6]. Furthermore, U and Th in the EZ have been migrating to the crust (these U and Th that have been moved to the surface have very little contribution in maintaining Earth's interior temperature) [5, 6, 14]. Therefore, the above-mentioned continuous warm-up in Earth's interior indicates that there exists at least one of the following two cases:

First, additional energy may be produced in Earth's interior through U, Th nuclear fission. Consequently, even though the total of U and Th in Earth's interior reduces with time, nuclear fission of these U and Th can produce enough additional energy to increase and maintain the Earth's interior temperature (see 3.2.2 and part 4).

Secondly, U and Th in the lower part of Earth (including lower mantle and the core) are relocated to the Earth's core, where the heat released from them is not easily dissipated, which make it possible to increase and maintain the temperature in Earth's interior.

In addition, Earth's cyclical expansions, volcanic activities, and other powerful cyclical geological activities imply that there were periods of rapid temperature increases during Earth's expansion phases. However, the decay of U and Th can only release heat gradually. Therefore, only when there are cyclical nuclear fission processes, such rapid increments in temperature could have occurred. Furthermore, only in the liquid outer core, U and Th could become enriched enough to trigger a spontaneous nuclear fission through gravity differentiation due to the large density difference between U and Th, and the surrounding liquid Fe (the major composition of the core) and Ni.

(5) The main geomagnetic field is dipole-symmetrical, so there must be a main thermal convection center in the outer core to produce this magnetic field. However, Earth's magnetic center deviates from its geographical center by 400km [37]. This indicates that the thermal convection center has deviated from the geographical center by 400km. This also implies that there is an asymmetrical heat source in the outer core. In addition, south and north geomagnetic poles are always migrating [37, 38]. Its migration speed ($0.004^{o}$~$0.00.7^{o}$/year) [38] is different from the differential rotation rate of the inner solid core relative to the solid Earth (several degrees faster /year [39]. Note: now it is considered $0.3^{o}$~$0.5^{o}$ faster/year, Zhang et al., Science 309, 2005, p1357). So the migration of main geomagnetic poles cannot be caused by the differential rotation of the inner core relative to the solid Earth, but could be related to the flow of heat source in the outer core. Only heat producing elements U and Th can be relatively concentrated in the outer core, and this concentrated center can easily and slowly migrate. As the Earth is rotating from west to east, the concentrated center of U and Th(**U, Th-rich center**), which have a much bigger density than the surrounding liquid Fe and Ni, will rotate behind the surrounding liquid Fe and Ni (and the whole Earth). This may make the U, Th-rich center migrate westwards relative to the whole Earth, which is consistent with the westward migration of main geomagnetic poles [2]. This U, Th-rich center may also cause the whole plate system to migrate westwards [40]. At the same time, the asymmetrical distribution of energy source in the outer core will lead to asymmetrical heat convection in the outer core, which may destroy the harmonic rotation between the mantle, outer core and inner core. This could produce differential rotation between Earth's different layers and allow for formation of the geomagnetic field [2, 36].

(6) Computer simulation studies [9, 10] show that if only the heat energy from radioactive elements is considered (according to the traditional radioactive element distribution model, U and Th mainly reside in the crust and upper mantle), there are a number of small-scale convections in the Earth's upper part, but almost no convection in the deep mantle. This is contradictory with the conclusion achieved with seismic tomographic mapping techniques that indicate whole mantle convection. If the heat source is supposedly from the earth's core, mantle convection is caused by only 2 ~ 3 super-plumes. This is consistent with global temperature profiles revealed by seismic tomographic techniques. Therefore, the energy to drive the mantle convection should come from the core [9, 10]. This requires heat producing elements in the core.

(7) A large number of studies have shown that rocks and their xenoliths from the deep mantle have extraordinarily a high amount of noble gases [41], and there are some patterns in the distribution of these rocks. Generally, the concentration of He and the $^{3}He/^{4}He$ ratio are very high in rocks from volcanic hot spots in the Pacific Ocean. Particularly, the rocks from Loihi and Samoa hot spots in Hawaii have the highest He concentration and $^{3}He/^{4}He$ ratio. Their average $^{3}He/^{4}He$ ratio is in the range of 11~24 Ra (Ra refers to this ratio of the atmosphere) [41~43]. This is followed by the volcanic rocks from Pacific ridges with an average value of ~8Ra. Generally, the volcanic rocks from Atlantic ridges are slightly lower with an average value of 7.7 Ra [41]. There is a consistent feature in these values from sea water of these two oceans. Of the two, the highest is in the Pacific Ocean, with $\Delta^{3}He$ reaching 11% in its deep-sea water. The Atlantic is systematically lower, and its $\Delta^{3}He$ is around 5.2% in its deep-sea water [41]. There also are positive abnormalities of Rn, $^{86}Kr$ [41], $^{129}Xe$, $^{136}Xe$ and $^{134}Xe$ in the xenoliths from hot spot volcanic rocks of Loihi and Samoa islands [41, 42]. The later three abnormal values can reach as high as 6%, 3.7% and 8%, respectively [42]. These noble gases are common in that they can be directly or indirectly produced by the decay or nuclear fission of U and Th [41]. Controversy surrounds the origin of these mantle noble gases [41, 44, 45]. Because it is difficult to give a reasonable explanation about the origin of these special noble gas isotopes, they are called “primitive gases” [41, 44] that were trapped in the lower mantle during Earth’s formation through accretion. However, this view cannot address the following questions [44, 45]: During Earth’s formation, the impacts by planetesimals or planetary embryos would produce high temperatures in the original Earth and might have led to formation of Magma Ocean. Therefore, primitive He would be released from rocks and escape from the original Earth under such extreme dynamic conditions because of the small mass of He. Therefore, this view mentioned above, in which high concentration of He and a high $^{3}He/^{4}He$ ratio come from “primitive gases” trapped in the lower mantle, cannot be supported by experiments, theory and observations [45]. If these noble gases are “primitive gases”, this demands that their source magmas have not been differentiated for the last 4.4Ga, which is controversial with evidences revealed by other geochemical indicator elements, and many isotope systems such as Sr, Pb etc [43, 45]. At the same time, the rocks containing these noble gases do not have primitive chondrite composition. Therefore, these noble gases cannot be referred to as the “primitive gases”. In addition, the mantle rocks discussed above also contain positive abnormality in $^{129}Xe$, which was the daughter element, through decaying of the extinct $^{129}I$ that has a half-life of 17 Ma. Traditionally, it was also assumed to be one of these “primitive gases”, but this view cannot answer many questions about its origin (41). If these noble gases are to be considered “primitive gases”, a relatively high concentration of $^{3}He$ and $^{129}Xe$ in these mantle rocks and their xenoliths also call for that their source rocks has been isolated from mantle differentiation and convection during the entire history of Earth’s evolution [41~44]. This requires layered mantle convection, which conflicts with evidence from other trace element geochemistry [44-45], and with the whole mantle convection model supported by seismic tomographic studies [9,10]. In order to solve these problems, Anderson proposed that a high He and $^{3}He/^{4}He$ ratio of the mantle rocks and lava may come from extraterrestrial materials that were carried down to the mantle with the subducting plates [45]. His major argument is that meteorites and other extraterrestrial materials have a very high He concentration and $^{3}He/^{4}He$ ratio. Secondly, this also helps to avoid the disadvantages of the "primitive gases" theory as discussed above [44, 45]. However, this assumption also has the following problems [44]: First, even if all of the $^{3}He$ from the extraterrestrial materials has been moved into the mantle by subduction, the total $^{3}He$ is still less than those escaped from the mantle, since $^{3}He$ can easily escape from the Earth, and does not readily enter the mantle. Secondly, there is no evidence to indicate that the $^{3}He$ from extraterrestrial materials can be kept in the sediments in the sea floor till it enters the subducting zones, since the $^{3}He/^{4}He$ ratio in the old sediments is very low [44].

In addition, seismic tomography and related computer simulations studies show that the super plume that reaches the outer core under the Pacific Ocean is supported by the heat from the core [9, 10]. Therefore, heat materials, such as magma, have been being released from the deep interior in this region because of heat from the core (Magma eruption in hot volcanic spots in Hawaii has lasted at least 50Ma [43]. If the 200Ma history of Pacific Ocean is considered, then these hot spot magmas may have erupted much longer); consequently, it is very difficult for materials in the lower mantle to maintain their original state and not undergo differentiation. Therefore, the hypothesis of "primitive gases" cannot be supported.

Because these noble gases can be produced directly or indirectly by the U, Th decay or fission [41], therefore, the fact that the volcanic rocks and their xenoliths in Pacific Ocean are rich in the noble gases as discussed above indicates a enrichment of U and Th in the outer core under the Pacific Ocean[30,31].

To sum up, there is a huge heat source in the outer core. Based on the foregoing U and Th geochemical principles, and the experimental study of Murrell et al. (1986), U and Th can enter the planetary cores and become one of the important energy sources [27]. Therefore, it is concluded that there is a lot of U and Th in the outer core. These U and Th have become relatively concentrated in certain parts of the outer core (forming a mobile U, Th-rich center). The heat convection center of the core deviates from the geographical center for approximately 400km.

### 3.1.2 The Position of U, Th-rich center in the outer core

The heat released from U and Th in the outer core will cause the overlying mantle and crust to be the hottest areas on the Earth. Anderson's studies indicate that the Pacific plate and its neighbouring regions are the hottest regions on the Earth [46]. The plumes located in central Pacific are the hottest, and they may be the plumes that originate from the deepest interiors of this planet. The hot volcanic spots from other areas, such as Iceland, and ridges in the Atlantic Ocean do not have heat sources with a depth larger than that of the asthenosphere, indicating that these hot spots and ridges are only cracks or cuts of the lithosphere [46]. The central Pacific continues to be the hottest area at 2300 km depth in the mantle [47]. Therefore, U, Th should have been enriched in the outer core under this area.

Secondly, the Pacific Ocean is the world's largest ocean basin, in which the Japanese ridges have produced the largest area in oceanic crust in the Earth's surface. At the same time, the plumes on Hawaiian and Polynesia islands have produced most of the hot spot volcanic rocks on the Earth [46]. This shows that the largest super-plume is located in the central Pacific Ocean, and is transferring heat and materials from the core and deep mantle to the surface [9, 10].

From 1300km, 1900km [48] down to 2300km [47], the cold mantle areas, the locations of the deep subduction zones, are gradually limited to the circum Pacific Ocean areas (Fig.1), indicating that only the super-plume and ridges in the Pacific Ocean can lead to the formation of the deep subduction zones down to the lower mantle and core-mantle boundary.

Along circum Pacific Ocean and the adjacent circum Indian Ocean, there lie huge and unique mountain ranges, including Cordillera mountain ranges with a length of ~ 15000km. This implies that the outer core under this area possesses the strongest power and heat source.

More importantly, in accordance with the 1980 South and North Magnetic Pole positions on the geo-globe (Surveying and Mapping Press, 1983), when the South and North magnetic poles are connected along with the shortest surface distance, it is found that this connection is right through the Central Pacific ocean area (Fig.1). The normal line at the mid-point of the magnetic axis that connects the South and North magnetic poles, also points to the direction of the central Pacific Ocean (Fig.2). Therefore, the geomagnetic center has deviated from the geographic center to the Pacific direction for 400km [39], which shows that the heat convection center that produces the Earth's magnetic field also deviate to the Pacific direction for 400km. Consequently, U and Th should just concentrate on the area in the outer core under the central Pacific Ocean.

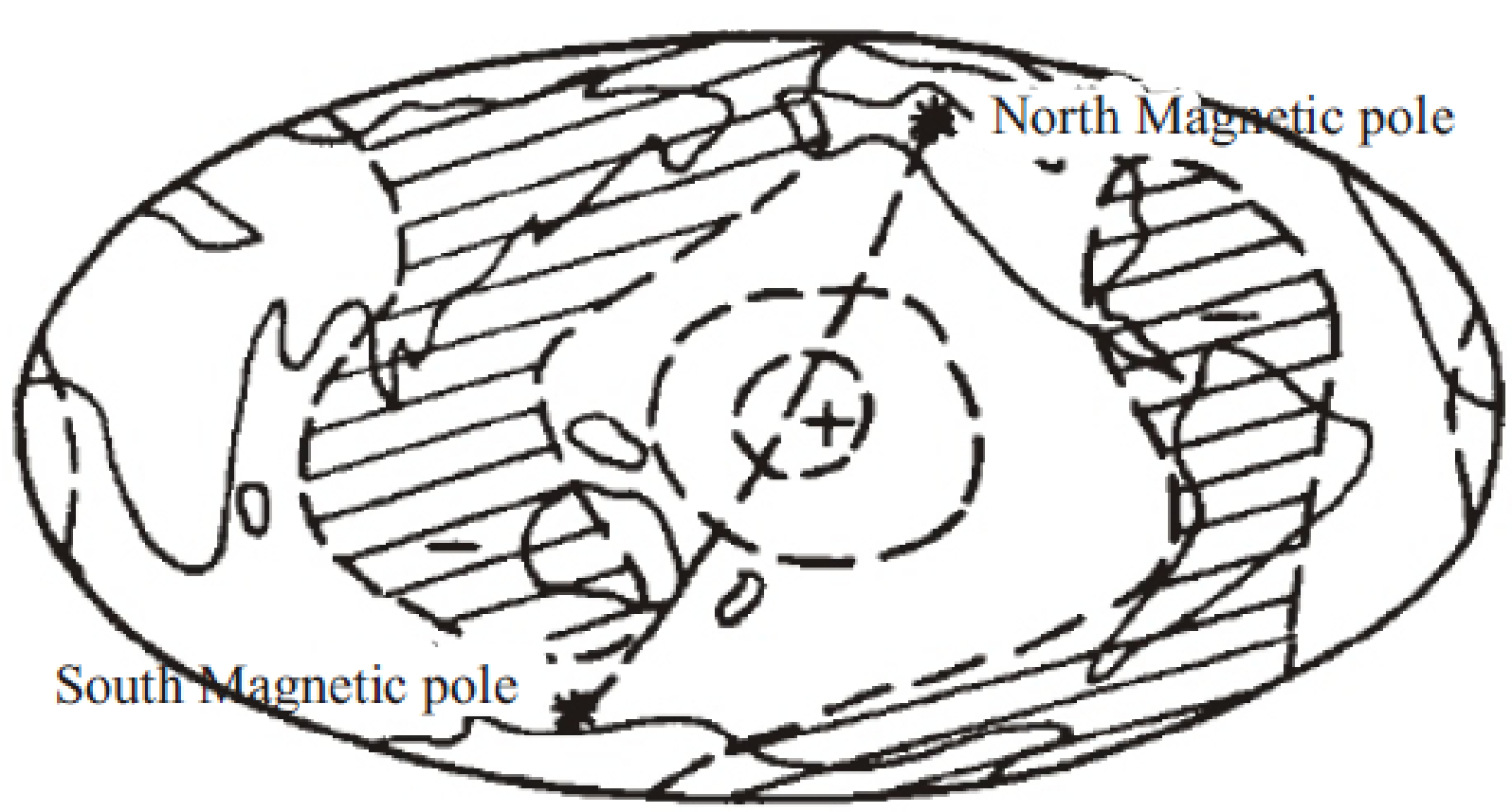

**Fig. 1 Temperature variation of the lower mantle at a depth of 2300 km (Based on the results from the seismic P-wave velocity of the lower mantle achieved by Anderson [46] and Carlo [47])**

**North and South magnetic poles and the shortest surface distance connected line between them are marked.**

**Where +: the hottest area; -: the coldest area**

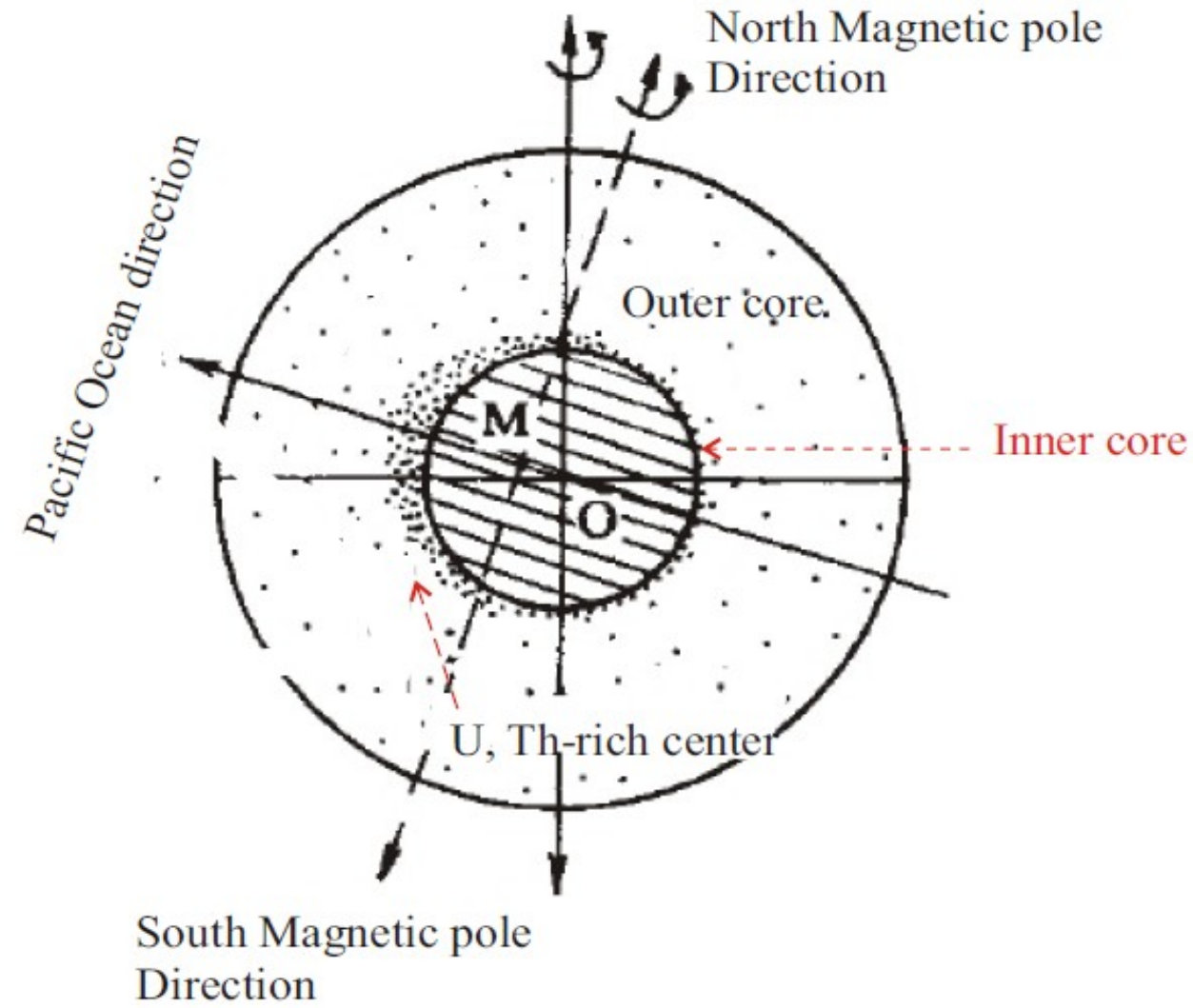

**Fig. 2 Distribution of U and Th in the Earth's outer core**
**Where O: Center of the Earth; M: center of the geomagnetic field and thermal convection in the core; the black dots present the relative concentration of U and Th.**

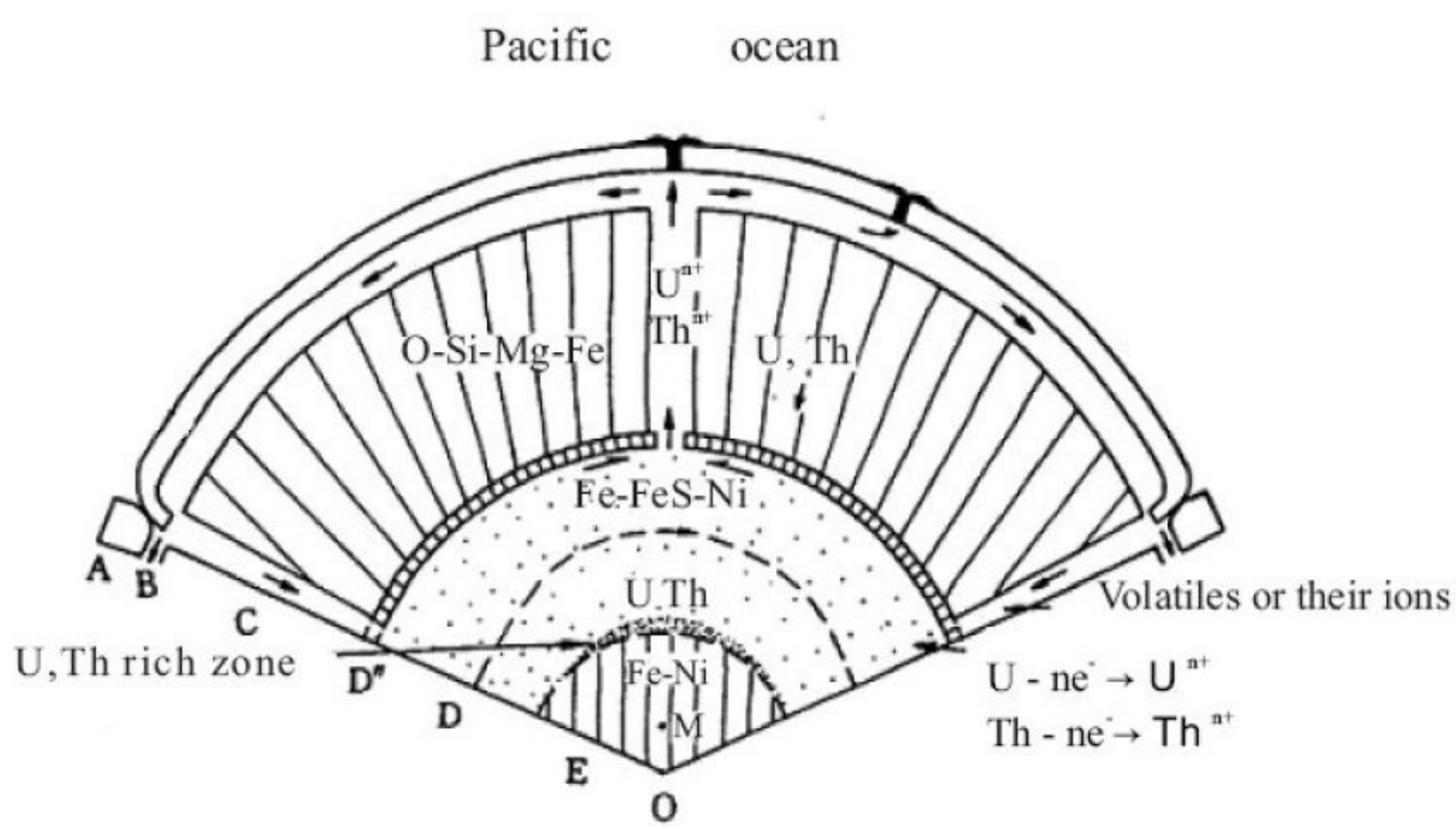

**Fig. 3 The distribution of U and Th in the outer core and its influence on the formation of deep mantle plumes and subducting zones in the Pacific Ocean.**

**Where A:-lithosphere; B: -U,Th enrichment zone or asthenosphere; C: - mantle; D'':- core-mantle boundary; D: - outer core; E: - inner core; O: - center of Earth; M: - center of geomagnetic field and thermal convection; the black dots represent the relative concentration of U and Th.**

As U and Th have a much greater density than Fe and Ni, this U, Th concentrated area should be closer to the bottom of the outer core under the centric Pacific area (Figs. 1~3).

### 3.1.3 The formation of an interior circulatory system

The nuclear energy released from U and Th in the U, Th-rich center will contribute to the formation of the super-plume in the Pacific Ocean. They bring up hot materials so that the Pacific region becomes Earth's hottest region; this also affects the neighbouring Indian Ocean region.

Because of the existence of abundant water and volatile components, U and Th in Earth's interior are partially oxidized and transformed into low valance components and compounds. They migrate to the asthenosphere position under the lithosphere first, and are enriched there. This leads to the formation of an enrichment zone of U and Th(EZ) [5, 6] in the position of the asthenosphere (Fig.3). This U, Th enrichment zone makes oceanic plates drift possible. The new oceanic crusts are continuously produced in the ridges, and then are pushed to both sides and drift over the asthenosphere (EZ), and finally subduct to the mantle. This leads to the formation of subduction zones. The subducting materials will carry a large number of crustal materials into the mantle and core. These cold crustal materials can cool the earth's core. At the same time, some of the oxidative volatile components or their ions from the subducting slabs can further be brought down to the core and oxidize the U and Th metals or their low valence compounds there; these U, Th metals or their low valence compounds are oxidized into relatively high valence U and Th components and oxides with a much smaller density, and then they migrate up together with the super-hot materials along the super-plumes into the asthenosphere [5, 6] (Fig.3). This constitutes a circulatory system for material movement in Earth's interior, and forms a relatively gentle way for Earth to release its interior heat, especially the heat sources, U and Th, which is the prerequisite for the origin and evolution of life [5, 6].

It was the circulatory system that led to the formation and development of the plate tectonics system. The energy to make this system work is the heat energy from the U and Th in the U, Th-rich centers in the outer core and in the U, Th enrichment zone in the asthenosphere. As mentioned above, water and oxidative volatile components, and their ions from the subducting crustal materials will partially enter the U, Th enrichment zone, and oxidize the U and Th there. This increases their mobility due to density decrement, and makes them migrate up into the crust, which makes U, Th content in the continental crust increase [5, 6, 14], but U, Th content in the U, Th enrichment zone decrease. In order to maintain the existence of an enrichment zone of U and Th (or asthenosphere), it is necessary to continuously add heat producing elements U and Th into this zone. The existence of water and oxidative volatile components in Earth make the U and Th in the outer core gradually move up to the enrichment zone of U and Th through the circulatory system discussed above.

Therefore, the existence of oxidative volatile components and water in Earth is the prerequisite to the formation and development of a plate tectonics system [5,6]. As shown in Fig.1, the temperature map at 2300 Km deep in the mantle clearly indicates the existence of the circum Pacific Ocean subduction zone (low temperature area) and Central Pacific super-plume (high-temperature zone) [47], supporting the existence of a circulatory system in Earth's interior.

Since the outer core has always been in a state of convection, the U,Th-rich center will also move slowly there, which will cause the super-plume to move slowly too. This is inconsistent with the traditional plume theory, but match with the results revealed by new seismological studies [46].

### 3.2 The relation between U, Th in the core and the superchrons of geomagnetic field

#### 3.2.1 The problems with current theories

Paleomagnetic studies have shown that in the last 350Ma, there have been two superchrons of geomagnetic field, that is, the Cretaceous period (124 ~ 83Ma) normal superchron, and Permian-Carboniferous (320 ~ 250Ma) reverse superchron. During these periods, the geomagnetic field weakened, to about 1/3~1/10 of the current value (note: now the field is considered much stronger (see refs. in Courtillot and Olson, EPSL 260, 2007, p495), and its reversal stops [2, 17~20]. Correspondingly, on the Earth's surface, sea-floor spreading, global climate change, plume activities, large magmatic activities and the formation of continental flood basalts are exceptionally prevalent and powerful [17~20]. At the same time, mass extinctions occurred, including the extinction of dinosaurs [17] (now it is considered that these strong geological events occurred 10-20 Ma after the end of the corresponding superchron(Courtillot and Olson, above). It is accepted that the magnetic field is generated through thermal convection in the outer core, and heat convection intensity in the outer core is directly proportional to the temperature gradient [17~20]. In general, temperature is gradually reduced from the inner core-outer core boundary to the core-mantle boundary [17-20].

Currently, there are two main views about the formation of the superchrons [20]. Courillot et al. suggested that when the core-mantle boundary layer D " thickens to the greatest thickness, it will thin and break. So the heat in the core will leak out from the core and promote the formation of plumes. The formation and activity of these plumes also intensify volcanic and geological activities, and mass extinction in the crust. The top of the core begins to cool down due to the loss of heat to the mantle, thus leading to a bigger difference in temperature between the top and bottom of the outer core, which reinforces thermal convection. When the thermal convection builds to a relatively high level, the Earth's magnetic field will weaken and stop its reversal [17]. However, Larson et al and Prevot et al. argued that that a strengthening of the heat convection will increases the energy of the Earth's dynamo, so that the Earth's magnetic field intensity and frequency of magnetic reversal will increase[18, 19]. This, in turn, is contradictory with the observed results that Earth's magnetic field weakened (about 1/3 of current value, note: new studies support a strong field during superchrons see Courtillot and Olson, above), and its reversal stops during the two periods of Cretaceous normal superchron and Permian-Carboniferous reverse superchron. As a result, the existing theories still have the following problems:

(1) These theories cannot reasonably explain the simultaneous phenomena: geomagnetic reversal stops, and magnetic field intensity increases.

(2) Despite the views of Courillot et al.[17], or of Larson et al.[18] and Prevot et al.[19], changes in the geomagnetic field is thought to have originated from the variation of heat convection intensity in the outer core. The change in heat convection intensity is attributed to the varying thickness of the core-mantle boundary layer D ", but they did not point out what causes the change in thickness of the CMB, D".

(3) As pointed out by Anderson, seismic studies have shown that in the bottom 200 km of the mantle, the seismic velocity suddenly increases. So it is a layer of materials with melt-resistant characteristics [16]. At the same time, all data show that from the mantle to the core, there is a sudden increase in temperature. So this layer also has the property of insulation. As a result, this layer of material(s) will destroy the so-called core-mantle coupling [18]. In addition, it has a strong tendency that the heat in the hotter lower mantle transmits to the colder upper mantle. So in turn, it is unfavourable for the colder mantle to provide heat energy [18,19] for the hotter core.

(4) During the superchron periods, not only the reversal frequency and the geomagnetic field intensity had large variations, but geological activities were exceptionally strong on the crust [2, 17~20], including a large number of biological extinctions [17]. At that time, Earth was strongly active from the core to the crust, indicating a process of energy explosion. However, there is no reasonable explanation provided on what the energy source is.

**3.2.2 Possible model**

As discussed above, there is a lot of U and Th in the outer core. They can increase the temperature of the outer core in two ways, and exert an impact on the thermal convection in the outer core: a) Over time, the heat released by U and Th decay will increase the temperature of the core. Correspondingly, the temperature of the lower mantle also increases due to the heat from the core. Consequently, the melting degree of the lower mantle also increases, which will lead to the U, Th in the lower mantle gradually to sink into the core and the total U and Th in the outer core increases. The temperature increment in the outer core will be speed up. b) When the U and Th in the outer core are enriched to a relatively high level, they will spontaneously produce nuclear fission in some areas. This will release a great lot of heat, and rapidly increase the outer core's temperature.

(1) Studies have shown that spontaneous nuclear reactions (nuclear fission) could occur in uranium mines in precambrian rocks. This kind of natural nuclear reaction could last up to one Ma [49]. Yang Tianxin [49] pointed out that there are three types of spontaneous neutron sources that can induce nuclear fission in nature, namely neutron can be released (1) when α particles, released from U, Th and its daughter elements, hit light elements such as $^{9}$Be, B, $^{18}$O, $^{8}$F etc., (2) when light elements, such as Be, $^{2}$H, are hit by the strong γ ray released by $^{208}$Tl, $^{214}$Bi, $^{212}$Bi and $^{234}$Pa, and (3) by the spontaneous fission of U. Many studies show that the core may have many light elements [25, 39]; a relatively high concentration of Tl and Bi in iron meteorites [50] indicates that the core may contain Tl, Bi etc. elements; $^{235}$U in natural uranium can be accounted for 0.72% (much higher in the Earth's early stages) [14, 41]. In fact, most of the neutron-producing sources are related to U and Th and their daughter/decay elements [51]. Therefore, the three neutron-producing sources could exist in the core.

$^{232}$Th and $^{238}$U can be transferred into fissionable isotopes $^{233}$U and $^{239}$Pu [51, 52] through absorbing neutrons (neutrons could come from, for instance, the spontaneous fission of $^{235}$U as discussed above). There is 0.72% fissionable isotope $^{235}$U in natural uranium [14, 41]. Therefore, when the U, Th concentrations reach a relatively high level, the concentration of fissionable isotopes will increase because of the fission induced by the neutrons from $^{235}$U. The neutron sources will be able to induce the formation of a

natural nuclear fission and corresponding nuclear reactions, and even value-added reactions [51].

The U and Th in the outer core are surrounded by liquid Fe-Ni alloy. Because of their much larger density [5, 6], these U and Th will easier sink to the bottom in the liquid outer core than in solid rocks (uranium ores) as mentioned above [49] and form relatively high concentrations. Therefore, they are likely to reach a concentration level to induce spontaneous nuclear fissions and form natural nuclear reactors, and produce a huge amount of heat.

(2) In addition, the volcanic rocks and their xenoliths from Hawaii and its surrounding areas of Pacific Ocean are rich in noble gas isotopes produced by the fission of U and Th. This also implies that nuclear fission reactions may occur in the outer core beneath [31].

(I) $^{3}He$ can be produced when Li, Be etc. light elements are hit by neutrons released by nuclear fission reactions, except coming from the so called “primitive gases” above.

Namely: 6Li + n (neutron) → $\alpha + {}^{3}H$; $^{3}H \rightarrow {}^{3}He + \beta$- and so on [14,41].

For instance, nuclear weapon experiments have caused $^{3}H$ to increase 500 times in rain water [53]. These $^{3}H$ will spontaneously decay into $^{3}He$ [41, 53]. Rocks that are rich in Li, U and Th are also rich in $^{3}He$ and He [14, 41]. This implies that the nuclear reaction above may exist in Earth’s interior.

Light elements such as lithium can enter the core through the circulatory system discussed above.

(II) Mantle rocks in the Pacific Ocean are rich in fission-producing $^{86}Kr$ and $^{136}Xe$ [41, 42]. This directly indicates of nuclear fission reactions in the Earth's interior, particularly in the outer core under the Pacific Ocean.

(III) The phenomenon that mantle noble gases are rich in $^{129}Xe$ also implies that U and Th nuclear fission reactions occur in the outer core. Also $^{129}Xe$ in nature can only be produced through U, Th nuclear fission or the decay of $^{129}I$ [41]. However, the original $^{129}I$ (half-life of 17Ma) has been extinction for 4.4 Ga [41]. The Hawaiian hot spot volcanoes have erupted at least 50 Ma. Therefore, the fact that the rocks from here are rich in $^{129}Xe$ [41, 42] indicates that there is nuclear fission in the outer core beneath or there is a source to continually produce $^{129}I$. Furthermore, only the nuclear fission of U and Th can continually produce $^{129}I$. For instance, the waste materials from nuclear reactors are rich in $^{129}I$ [53]. A positive correlation between $^{129}Xe$ and fission-genetic $^{136}Xe$ in these rocks [54] also indicates that $^{129}Xe$ (and $^{129}I$) comes from the nuclear fissions of U and Th. The $^{86}Kr/^{136}Xe$ ratio in the atmosphere indicates that the parent elements of this kind of nuclear fission are a mixture of $^{238}U$, $^{235}U$ and other isotopes [41].

(3) Planetary luminosity and mass relation studies shows that the nuclear reactions in the core are the main energy source of Earth’s evolution [28]. As the Earth’s core does not have the required high temperature of over $1x10^{9}$ $^{o}C$ for nuclear fusion reaction [29]. Therefore, this kind of nuclear reaction may be nuclear fission reactions of U and Th. The decay and nuclear fission of U and Th allow the Pacific Ocean to become a positive anomaly center of He, $^{3}He$, $^{3}He/^{4}He$ ratio, $^{129}Xe$ and $^{136}Xe$, since they can be directly or indirectly produced through the decay and nuclear fission of U and Th [41].

The two processes, particularly U, Th nuclear fission in the outer core could lead the outer core to boil or some of its light elements to boil, and solid Fe-Ni inner core to be molten and shrunken. Boiling will make the temperature gradient decrease in the outer core. Under such a condition, the thermal convection in the outer core will be stopped or weakened, which will suddenly weaken the geomagnetic field or make it disappear (see new explanation in part 4). At the same time, a shrunken inner core may result in the thermal convection disorders in the outer core, which will weaken the geomagnetic field, and make the geomagnetic field reversal stop. Many studies indicate that the existence of the solid inner core may play an important role in producing the geomagnetic field. Venus does not have a magnetic field, which may be because it does not have a solid inner core [55].

Tremendous heat generated by the nuclear fission events would cause the heat and heat materials (including light elements and oxidative U, Th compounds and oxides) to break through the core-mantle boundary layer D'', and migrate up through super-plume to the surface. This may result in strong crustal movement, and trigger the formation of the ocean, and mass extinctions. As the U and Th are relatively concentrated in the outer core under the Pacific Ocean, therefore, the formation of the Pacific Ocean may be related to the above-mentioned heat events.

After experiencing a long period of the above-mentioned nuclear fission events, the concentration of U and Th in the outer core decreases. The core will cool down after a long period of time. When the core temperature is cooled down to a relatively low level, the inner core will re-grow; thermal convection restarts; and the geomagnetic field strengthens gradually. During Mesozoic superchron, the intensity of the dipole geomagnetic field was reduced rapidly and then slowly strengthened [19]. This trend can be explained with the process described above.

In the process of the above heat events, a high mantle temperature will cause the U and Th in the mantle rocks to sink further to the core. A long period is needed for U, Th to reach the bottom of the outer core. A new superchron of geomagnetic field could occur again, when U and Th in the core increases to a relatively high concentration level to induce new nuclear fission and shrink the inner core. These events also could be the cause for Earth's cyclical expansions. It might also be the important trigger for the cyclical characteristics in geological and volcanic activities [3, 40].

## 4. Notes and an improved model

This is my paper published in Geological Review 45 (Sup.), 1999, 82-92.

After reading more literatures in English, I realized that there are other hypotheses on U nuclear fission in the Earth's interior. These include Herndon's U nuclear fission in the innermost inner core (J. Geomagn. Geoelectr. 45, 1993, p423), and Meijer et al.'s U nuclear fission in the core mantle boundary (**CMB**) (SAJS104, 2008, p111). However, I do not think it is feasible to consider U in the innermost inner core or inside the solid inner core. First, if U is in the inner core, the heat released from the decay or nuclear fission of U and Th will destroy the solid state of the solid inner, and make it disappear, which is not consistent with seismic observations that support a diameter of ~2550 km solid inner core. Although U has much larger density than that of Fe (and Ni), the major constituent of the core, gravity in the range of the inner core is only 0~40% of it in the surface (Bullen, Tran. Proc. Roy. Soc. New Zeal. 69, 1940, p188). Therefore, the large density of U cannot alone make it enter the inner core. On the contrary, U will stay in the liquid outer core due to its lower melting point than that of Fe (see following), and its low concentration in Earth's composition. Similarly, if U is enriched in the CMB, the heat released from it could also molten and destroy the CMB. According to our experimental results (Bao et al., 2008 http://geonu.snolab.ca/talks/bao.ppt), the oxygen fugacity at the CMB may be lower than -12 ΔIW. Under such highly reducing conditions, the partitioning coefficient $D_U$ (U concentration in metal/U concentration in silicate) will be > 0.09~0.17(Bao et al., 2008, above). Gravity effect at the CMB is 1.05 times larger than it in the surface (Bullen, 1940, above). Therefore, even a small $D_U$ can make the U in the CMB gradually sink to the liquid outer core. Also, current geoneutrino detecting techniques are unable to determine the direction and distance that an anti-neutrino travels from (private communication with Dr. Enomoto in 2008 geoneutrino meeting: Enomoto, 2008: http://geonu.snolab.ca/talks/enomoto.pdf), although they can be used to identify whether an anti-neutrino comes from U and Th, or from K atoms according to its energy level (Araki et al., Nature 436, 2005 p499). Currently, we are unable to identify whether a neutrino comes from U and Th in the mantle, crust or core, since current calculated results based on geoneutrino detecting techniques completely depends on author's favourable geochemical models (Bao et al., 2008 above). In addition, the contribution to the total heat flux from secular heat is limited to the first billion years after Earth's formation based on the studies of Van Den Berg and Yuen (EPSL 199, 2002 p403) and Van Den Berg et al.(PEPI 129, 2002 p359). Therefore, U and Th may still be the main energy source in Earth's deep interior.

The following is an improved model:

As discussed above, the force of gravity of U and Th is not able to make them enter the center of the core under the physical and chemical conditions of the core. On the contrary, Coriolis force (a combination of the centrifugal force and force produced by Earth's rotation) and the attractive force between atoms of U and Th may make them concentrate in area close to the equator from the middle to the bottom of the liquid outer core (**lower outer core**). In the range of the current lower outer core, gravity is only 70% ~ 40% of that on the Earth's surface. In addition, the melting point of U (1132 $^{o}$C) at the ambient pressure is 403 $^{o}$C lower that of Fe (1535$^{o}$C). This implies that when Fe crystallizes into solid crystal, U still stays in the melting liquid. At the same time, the radii of U and Th (156 and 179 picometer (pm), respectively) are much larger than that of the surrounding Fe (126 pm). Therefore, U and Th will stay in the loose liquid outer core with the liquid Fe rather than in the tight solid inner core with solid Fe crystal. In addition, the density of U and Th is 2.42 and 2.25 times respectively larger than that of the surrounding Fe at ambient conditions; therefore, Coriolis force (2ρ **Ω** u, here ρ-density, **Ω**-rotation rate of the Earth, u- the convection velocity in the outer core. Jones, Treatise on Geophysics, V8, 2006) of U and Th will be much larger than that of the surrounding Fe. Large Coriolis force of U and Th will make them stay in the areas close to the equatorial position in the lower outer core. Note that U and Th in the U, Th-rich centers could be U, Th metal or their low valance oxides and compounds [6]. If they exist as U, Th low valance oxides and compounds, the density difference between U, Th low valance oxides and compounds, and the surrounding Fe decreases (refer to [6]). Correspondingly, their Coriolis force decreases. However, the negative buoyancy force of U, Th low valance oxides and compounds also decreases; namely, the potential of U and Th entering

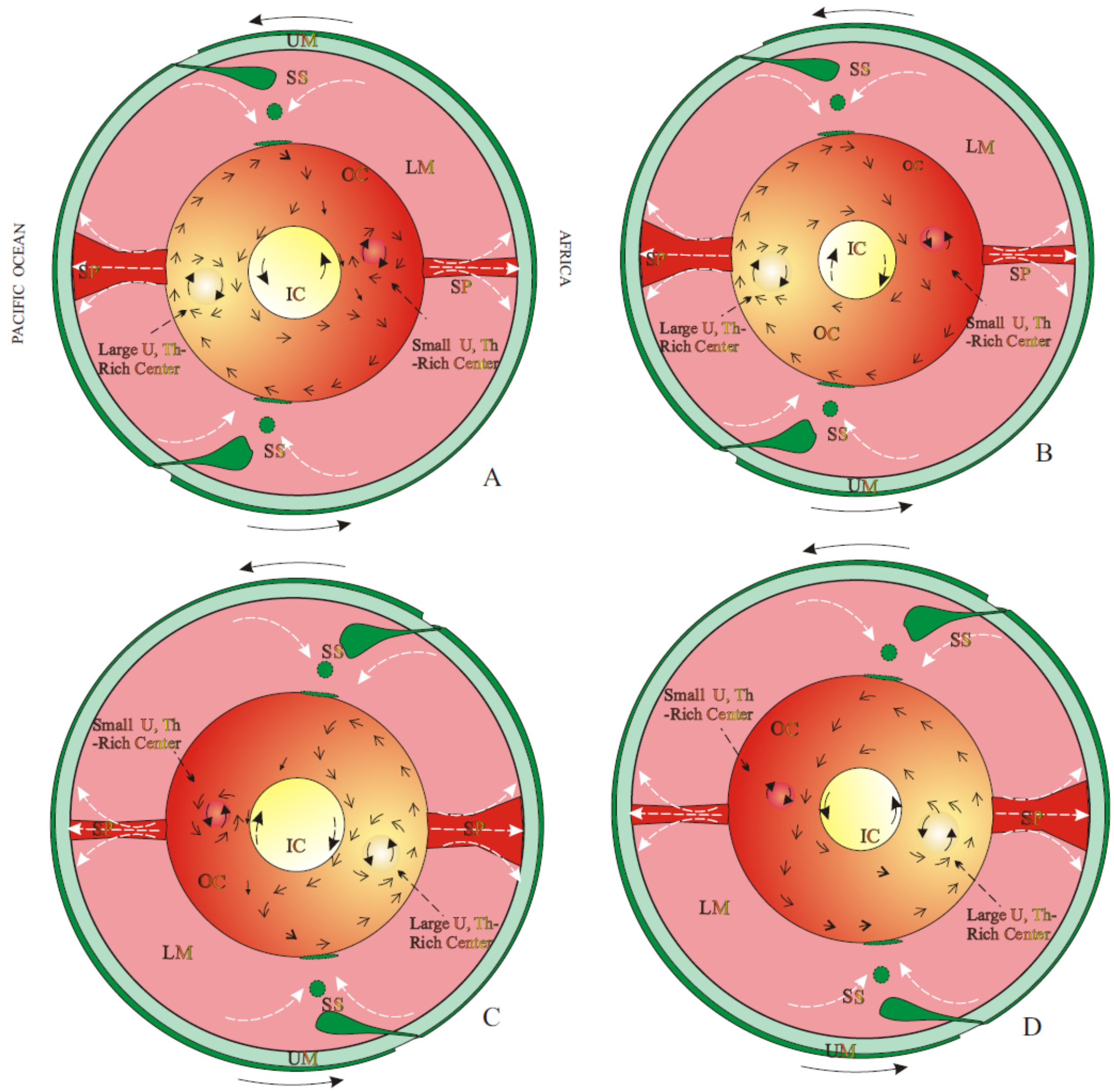

**Fig. 4 Cross section through equatorial plane viewed from the North Pole showing the U, Th distribution in the outer core and its influence on the heat convection in the core, and the mantle convection.**

**Where UM: upper mantle; SS: subducted slab; LM: lower mantle; SP: super-plume; OC: outer core; IC: inner core. White, yellow and red represent the relative high, medium and low temperatures, respectively in the core. The Earth rotates counterclockwise (eastwards) when it is viewed from the North Pole. Dotted line arrows represent that the inner core rotates at a speed of slower-than-Earth, namely it rotates westwards (clockwise in Fig.4) relative to the mantle. The large U, Th-rich center may be closer to the inner core. Here I drew it in the mid-outer core position for the convenience to draw the convection structure around it. The U and Th in the U, Th-rich centers could be U, Th metal or their low valance oxides and compounds [6].**

**A) Current Earth: there is a large U, Th-rich center in the lower outer core under the Pacific Ocean, and a small U, Th-rich center in the lower outer core under the Africa (A detailed explanation about their formation can be found in the text). Coriolis force (a combination of the centrifugal force and force produced by Earth's rotation) and super-rotation of the solid inner core make the large U, Th-rich center rotate clockwise. The large U, Th-rich center and its rotation control the heat convection pattern of the outer core and the temperature gradient in**

**the inner core, namely the temperature of the western hemisphere is higher than of the eastern hemisphere.**

**B) When the large U, Th-rich center attracts more U and Th to it, the heat convection triggered by the heat released by U and Th in this center strengthens. This will force the rotation of the small U, Th-rich center to change from clockwise to the counter-clockwise direction, and also the inner core to rotate from faster-than-Earth to slower-than-Earth (dotted line arrows represent a westward rotation of the inner core relative to the mantle). When strong nuclear fission occurs in the large U, Th-rich center, the inner core could be molten and shrunken, and consequently, a normal superchron could occur.**

**C) When strong nuclear fission occurs in the old large U, Th-rich center as indicated in Fig.4B, it will lose its U and Th and evolve into a small U, Th-rich center due to its high temperature and strong convection. At the same time, the small U, Th-rich center could develop into a new large U, Th-rich center. When the new large U, Th-rich center and its heat convection controls the whole core, a geomagnetic reversal will occur.**

**D) When strong nuclear fission occurs in the new large U, Th-rich center, the high temperature could make the inner core molten and shrink again, and a reverse polarity geomagnetic superchron could occur.**

the Earth's inner core decreases. Therefore, U and Th still stay in the areas close to the equatorial position in the lower outer core. Furthermore, U and Th may not be able to form metal or other crystals under the physical and chemical conditions of the core. One of the reasons may be that the heat released by their decay or nuclear fission will prevent them from reaching a high concentration, therefore, they are scattered in the liquid outer core. The attractive force among the atoms of U and Th will make them form enrichment centers in the area close to the equator in the lower outer core, because of their much larger density than the surrounding Fe.

At the beginning, several U, Th-rich centers might have developed. A combination of rotation of Earth, the super rotation of the inner core (Zhang et al, Science 309, 2005, p1357) and the Coriolis force of the U, Th atoms, and the heat released from these U and Th in the centers makes these U, Th-rich centers rotate and create convection around themselves respectively in the outer core. Furthermore, the heat convection also creates chance for the relative large centers to swallow the relative small centers, since the larger U, Th-rich centers can attract more U and Th with their stronger attractive force (or stronger affinity among the atoms of U and Th). At last, only one of the large centers develops into a single large U, Th-rich center as indicated in left side of the outer core in Fig.4A. However, in the opposite side of this large U, Th-rich center past the solid inner core, a small U, Th-rich center could survive due to the separation from the large U, Th-rich center by the solid inner core. The heat released from the large U, Th-rich center may have triggered the formation of a larger super-plume in the Pacific Ocean area, and the small U, Th-rich center has led the formation of the smaller super-plume in the Africa area as indicated in Fig.4A. Note that the positions of the super-plumes in both the Pacific ocean and Africa deviate from the U, Th-rich centers westwards due to the directional effect of the heat convection on the heat transfer between the super-plumes and the U, Th-rich centers as indicated in Fig.4A.

Similarly, the heat released from the large U, Th-rich centers also make the west hemisphere of the solid inner core much hotter than the East hemisphere. Note that the hottest position in the solid inner core deviates from the large U, Th-rich center eastwards due to the directional effect of the heat convection on the heat transfer between them as indicated in Fig.4A. It is considered that the heat convection is dominated by the large U,

Th-rich center here as indicated in Fig.4A. Therefore, the temperature distribution of the solid inner core is completely controlled by the heat convection of the large U, Th-rich center. Namely, the temperature of the inner core is higher in the western hemisphere and lower in the eastern hemisphere since the former is closer to the large U, Th-rich center and the latter is farer away from it. This causes the seismic elastic anisotropy to be larger in the west hemisphere than in the east hemisphere of the solid inner core as revealed by Sun and Song (EPSL 269, 2008, p56), since the relatively high temperature phase of fcc- or bcc-iron has larger elastic anisotropy than the relatively low temperature phase of hcp-Fe (Dubrovinsky et al., Science 316, 2007, p1880; Belonoshko et al, Science 319, 2008, p797; Kuwayama et al., EPSL 273,2008, p379).

However, when the concentration of U and Th is too high (for instance, > a few percentages) or nuclear fission occurs in the large U, Th-rich center, the huge amount of heat from the center will make the convection very strong. This will change the convection structure in the outer core as indicated in Fig.4B, and slow down the inner core super rotation, and finally make it rotate slower than the solid Earth; namely the inner core will rotate westwards(clockwise in Fig.4) relatively to the solid mantle as shown by the dotted line arrowheads in Fig.4B. At the same time, the heat from the U, Th nuclear fission will cause the inner core to become molten and shrunken, which will cause the geomagnetic field stable and stop reversing; namely, it leads to the formation of a normal polarity superchron (Coe and Glatzmaior, Geophys. J. Int. 166, 2006, p97, and also Bao et al., 2008 above). Also the strong convection will make the geomagnetic field strengthen during the superchron period, and lead to a much-stronger-than-average intensity as summarized by Courtillot and Olson (EPSL 260(2007), 495). In addition, the high temperature and strong convection will lead the large U, Th-rich center to disperse U and Th from this center, and it gradually loses U and Th to the opposite small U, Th-rich center. When the U, Th total in the small center can compete with or is larger than that of the large U, Th-rich center, the small U, Th-rich center will develop into a new large U, Th-rich center due to its relatively low temperature and weak convection as indicated in Fig.4C. Because the effect from the slower-than-solid Earth rotation rate of the solid inner core and the strong clockwise convection from the old large U, Th-rich center (Fig.4B), the convection direction of the small U, Th-rich center (namely the current new large U, Th-rich center in Fig.4C) will be changed from original clockwise to counter-clockwise direction as indicated in Fig.4 from A, B to C. After it develops into a new large U, Th-rich center as indicated in the right side of Fig.4C, its convection model will dominate the convection structure in the outer core, and this will make a geomagnetic reversal occur. However, on the opposite side of the new large U, Th-rich center, a new small U, Th-rich center (on the left side of Fig.4C) can be developed directly from the weakening old large U, Th-rich center due to the “protection” of the solid inner core. When the concentration of U and Th in the new large U, Th-rich center increase to a relatively high level, a new nuclear fission event could occur again as indicated in Fig.4D. This will lead to the next geomagnetic reversal, and if this heat event can molten the inner core and make it shrink, a new reverse polarity superchron of geomagnetic field could occur too.

In summary, the large and the small U, Th centers always compete and switch with each other. When a switch is successful, a geomagnetic reversal occurs. When strong nuclear fission occurs in the large U, Th-center and the inner core is molten and shrunken, a geomagnetic supperchron could occur. The evolution of Earth could go along with the order of ABCDABCD… when strong nuclear fission occurs in the large U,Th-rich center, which may trigger a corresponding geomagnetic superchron. Or, its evolution goes in the order of ACAC… when only weak nuclear fission occurs in the large U, Th-rich center, which may only trigger geomagnetic reversals. More realistically, the Earth may evolve in a mixed order of the two sequences above. However, the behaviour of the geomagnetic field may become very different (for instance, it may not switch its polarity in the future), when the concentration of U and Th in the outer core drops to a relatively low level, and the large U, Th-rich center is too weak to influence the convection of the whole core.

As pointed out by Jones (2006, above), the biggest progress in understanding the dynamo that generates the geomagnetic field has been made through computational simulations. Currently, Earth's magnetic field only be modeled with grossly simplified parameters (Christensen, Nature 454, 2008, p1058). For instance, the Ekman number E (the ratio of viscous to Coriolis forces, Walker et al., Geophys. Astrophys. Fluid Dynamics 88, 1998, p261), a key parameter of computational simulations, is only around $10^{-15}$ in the Earth's core ( Walkler et al., above; Christensen, above). However, current simulations only can adopt an E value of $\sim10^{-4} – 10^{-6}$(see the simulations of Jones, 2006 above, Aubert et al. Nature 454, 2008, p758, etc.). When a more realistic E value of $\sim10^{-7}$ was used, the simulated magnetic field became very different from the Earth's field, and was not predominantly dipolar (Kageyama et al. Nature 454, 2008; Christensen, above). This unrealistic and opposing result strongly implies that current simulation studies may have overlooked key factors, such as energy source, U and Th, and their nuclear fission in the outer core, etc. Other important parameters in computational simulations, such as magnetic Prandtl number (The ratio of viscosity to electrical resistivity) of the core is about $10^{-6}$, but it is usually set to 1 in current simulations (Christensen, above). These suggest the need to break through the traditional way of thinking about the core.

Secondly, the geomagnetic superchrons were usually accompanied by strong geologic activities and mass extinctions in Earth's crust in a 10~20 Ma lag (Courtillot and Olson, above). This implies that there is a strong energy explosion from the outer core during a superchron period, and cannot be explained by a secular cooling process of the core. In addition, paleomagnetic records indicate a long-term asymmetric saw-toothed pattern of geomagnetic field intensity during geomagnetic reversals. When a reversal was about to occur, the geomagnetic field usually strengthened to maximum rapidly, and then dropped slowly following a reversal (Hoffman, EOS 76, July 18, 1995 p289; Valet et al., Nature 435, 2005, p802). This intensity variation model, including geomagnetic reversals, cannot be reasonably explained by traditional secular cooling process of the core, but can be well explained by the model discussed above. Namely, when the U, Th total in a small U, Th-rich center surpasses that of the old large U, Th-rich center (in the left side of Fig.4B), this small center (in the right side of Fig.4B) will quickly attract more U and Th from the weakening old U, Th-rich center, and evolve into a new large U, Th-rich center as indicated in the right side of Fig.4C. This will cause a geomagnetic reversal. The convection of this new large U, Th-rich center will strengthen to maximum quickly, but be weakened gradually. This is because when the U, Th concentration in the new large U, Th-rich center reaches a high level, strong convection driven by the high heat produced by U, Th decay or fission will disperse U and Th from this center. Consequently, after the energy level in the new center reaches maximum, it will be weakened gradually from losing U and Th. Correspondingly, this will cause the magnetic intensity to increase to maximum rapidly, and then decrease gradually. If Earth's evolution follows the steps of ABCDA… or ACAC…, or a mixed order of the two, an asymmetric saw-toothed pattern of geomagnetic field intensity described above will occur. Furthermore, the supercontinent cycle (Wilson cycle) (the aggregation and dispersal of supercontinents) might have happened many times in the last 3.0 Ga (Gurnis, Nature 332, 1988, p695). This can also be reasonably explained by the evolution process of ABCDA…, or ACAC…, or a mixed order of the two in Fig.4.

In conclusion, to better understand Earth's geomagnetic field, it is necessary to add more considerations to current computational simulations. As we know, the Sun's magnetic field reverses its poles about every 11 years (Zhang and Schubert, Rep. Prog. Phys. 69, 2006, p1581). Some researchers have compared the Earth's dynamo to Sun's dynamo (Zhang and Schubert, above). However, Sun's dynamo is supported by nuclear fusion reaction of H, He etc. in its core. Therefore, we could also consider that Earth's dynamo is supported by nuclear fission reaction of U and Th in the outer core based on the discussion above. Fusion and fission reactions are just two opposite nuclear reactions in nature. It may be one of the universe symmetries, although it is not a perfect one. Therefore, nuclear fission reactions may also be common in some planetary cores, just as nuclear fusion reactions are common in the core of stars.

**Acknowledgements**

Many thanks for the support from my supervisor Dr. R.A Secco, and CFI.